\documentclass{ws-rv9x6}  
\usepackage{graphicx,hyperref}
\usepackage[sectionbib]{chapterbib}
\newcommand{\eq}[1]{Eq.~(\ref{#1})}
\newcommand{\fig}[1]{Fig.~\ref{#1}}
\newcommand{\dotv}{\mbox{\boldmath\(\cdot\)}}
\newcommand{\ddotv}{\mbox{\boldmath\(:\)}}
\newcommand{\cross}{\mbox{\boldmath\(\times\)} }
\newcommand{\grad}{\mbox{\boldmath\(\nabla\)} }

\newcommand{\curl}{\mbox{\boldmath\(\nabla\times\)} }
\newcommand{\const}{{\mathrm{const}}}
\newcommand{\Bvec}{{\mathbf{B}}}
\newcommand{\Evec}{{\mathbf{E}}}
\newcommand{\Kvec}{{\mathbf{K}}}
\newcommand{\kvec}{{\mathbf{k}}}
\newcommand{\rvec}{{\mathbf{r}}}
\newcommand{\vvec}{{\mathbf{v}}}
\newcommand{\xhat}{\hat{\mathbf{x}}}
\newcommand{\yhat}{\hat{\mathbf{y}}}
\newcommand{\zhat}{\hat{\mathbf{z}}}

\newcommand{\half}{{\textstyle{\mathrm{\frac{1}{2}}}}}

\newcommand{\phibar}{\overline{\varphi}}
\newcommand{\phitil}{\tilde{\varphi}}
\newcommand{\psibar}{\overline{\psi}}
\newcommand{\psitil}{\tilde{\psi}}
\newcommand{\Pbar}{\overline{P}}
\newcommand{\Ptil}{\widetilde{P}}
\newcommand{\rhos}{\rho_{\rm s}}

\renewcommand{\ij}{{\rm i}}
\newcommand{\D}{{\rm d}}

\begin{document}

\setcounter{chapter}{0}

\chapter{ZONAL FLOW GENERATION BY MODULATIONAL INSTABILITY}

\markboth{R.L. Dewar \& R.F. Abdullatif}
{Zonal flow generation by modulational instability}

\author{R.~L. Dewar and R.~F. Abdullatif}

\address{Department of Theoretical Physics\\
Research School of Physical Sciences \& Engineering\\
The Australian National University\\
Canberra ACT 0200, Australia\\
E-mail: robert.dewar@anu.edu.au}

\begin{abstract}
		 This paper gives a pedagogic review of the envelope formalism for
		 excitation of zonal flows by nonlinear interactions of plasma
		 drift waves or Rossby waves, described equivalently by the
		 Hasega\-wa--Mima (HM) equation or the quasigeostrophic barotropic
		 potential vorticity equation, respectively.  In the plasma case a
		 modified form of the HM equation, which takes into account
		 suppression of the magnetic-surface-averaged electron density
		 response by a small amount of rotational transform, is also
		 analyzed.  Excitation of zonal mean flow by a modulated wave
		 train is particularly strong in the modified HM case.  A local
		 dispersion relation for a coherent wave train is calculated by
		 linearizing about a background mean flow and used to find the
		 nonlinear frequency shift by inserting the nonlinearly excited
		 mean flow.  Using the generic nonlinear Schr\"{o}dinger equation
		 about a uniform carrier wave, the criterion for instability of
		 small modulations of the wave train is found, as is the maximum
		 growth rate and phase velocity of the modulations and zonal
		 flows, in both the modified and unmodified cases.
\end{abstract}

\section{Introduction}
\label{sec:intro}

As of January 1, 2006, Wikipedia\cite{wikipedia06} introduces the
term ``zonal flow'' thus:
\begin{quote}
    Fluid flow is often decomposed into mean and deviation from the
    mean, where the averaging can be done in either space or time,
    thus the mean flow is the field of means for all individual grid
    points.  In the atmospheric sciences, the mean flow is taken to be
    the purely \emph{zonal flow} of the atmosphere which is driven by
    the temperature contrast between equator and the poles.

    In geography, geophysics, and meteorology, \emph{zonal} usually
    means `along a latitude circle', i.e. `in the east-west
    direction'.  In atmospheric sciences the zonal coordinate is
    denoted by $x$, and the zonal wind speed by $u$.
\end{quote}

\begin{figure}[h!]
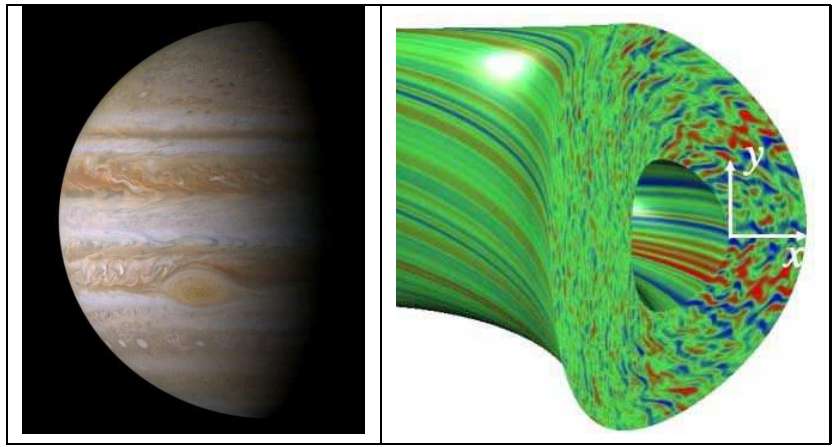

    \centering
    \begin{tabular}{|c|c|}
        \hline
 	\psfig{file=NASA_Jupiter_PIA04866_modest.epsf,scale=0.25}
&     
 	\psfig{file=Candy-torus-xy.epsf,scale=0.45}
  \\
        \hline
    \end{tabular}
    \caption{Left panel shows NASA image PIA04866: Cassini Jupiter
    Portrait, a mosaic of 27 images taken in December 2000 by the
    Cassini spacecraft.  Right panel shows a simulation of plasma
    potential fluctuations in a tokamak cut at a fixed toroidal angle
    as produced by the GYRO code (courtesy Jeff Candy
    \url{http://fusion.gat.com/theory/pmp/}). Note that, in the 
     plasma case, zonal flows are in the $y$-direction when slab 
     geometry is used.}
    \label{fig:JupitervsTokamak}
\end{figure}

The above definition omits the further qualification that zonal flows
are restricted to bands (zones) of latitude.  This is most clearly
seen in the banded cloud patterns on Jupiter (see left panel
of \fig{fig:JupitervsTokamak}) where the magnitude, and even sign, of
the zonal flows varies with latitude in a quasiperiodic fashion. (On 
Earth, topographic variations like mountain ranges disrupt the zonal 
symmetry.)

The term ``zonal flow'' has also recently come to be much used in
toroidal magnetic confinement plasma physics (see e.g. the review of
Diamond \emph{et al.}\cite{diamond-itoh-itoh-hahm05}) to refer to a
mean poloidal flow with strong variation in minor radius.  The sheared
nature of this flow is thought to have the strongly beneficial effect
of reducing radial transport by suppressing turbulence, thus improving
the confinement of heat required to achieve fusion conditions.

The use of the same phrase ``zonal flow'' in the context of both
geophysics and magnetic plasma confinement is no coincidence, as the
existence of strong analogies between these fields has become well
recognized.\cite{horton97}  In this paper we work in the plasma
context, but point out the relation to the geophysical context when
appropriate.

The right panel of \fig{fig:JupitervsTokamak} depicts a section of a
simulated tokamak, showing turbulence excited by gradients in
temperature and density (the plasma being hotter and denser in the
middle section, which has been cut out to aid in simulation and
visualization).  The radial coherence length of the drift-wave eddies
has been reduced by spontaneously excited zonal flows.  The magnetic
field is predominantly in the toroidal direction, but there is some
component in the poloidal direction so that the magnetic field lines
wind helically around the torus, mapping out nested toroidal
\emph{magnetic surfaces} that confine the magnetic field in
topologically toroidal magnetic flux tubes.  The helical nature of the
magnetic field lines can be seen in the figure from the fact that the
turbulent eddies have their cores essentially aligned with the
magnetic field, making the turbulence quasi-two-dimensional despite
the manifestly three-dimensional nature of the tokamak.

A generalized polar representation would clearly be most appropriate
for representing the cross-sectional plane of the torus, but, for the
purpose of gaining physical insight with a minimum of formalism in
this paper we use \emph{slab geometry}.  That is, the toroidal
magnetic surfaces are imagined as flattened into planes, so that
Cartesian coordinates, $x$, $y$, $z$, can be used, with $y$ and $z$
replacing the poloidal and toroidal angles, respectively, and $x$ the
minor radius.  The $x$ and $y$ directions are indicated in the right
panel of \fig{fig:JupitervsTokamak}.  The slab approximation is the
analogue of the $\beta$-plane approximation in geophysics, but note
the axis convention is opposite to that used in geophysics, with $y$
now the zonal direction.

Modulational instability of drift waves (the analogue of planetary
Rossby waves) is a strong
candidate\cite{diamond_etal98,smolyakov-diamond-shevchenko00,chen-lin-white00,guzdar-kleva-chen01,lashmored-mccarthy-thyagaraja01,champeaux-diamond01}
for generating these zonal flows through a feedback mechanism, in
which modulations of the wave envelope excite zonal flows through a
nonlinear mechanism (Reynolds stress) and the zonal flows enhance the
modulation through a self-focusing mechanism.

It is the aim of this paper to elucidate this theory in a pedagogic
way using as simple a plasma description as possible, namely the
one-field Hasegawa--Mima
equation.\cite{hasegawa-mima78,hasegawa-maclennan-kodama79}  This
provides a simple theoretical starting point for describing the
nonlinear interaction of drift waves and zonal
flows.\cite{hasegawa-maclennan-kodama79,smolyakov-diamond-shevchenko00}
The same equation also describes Rossby wave turbulence in planetary
flows in the quasigeostrophic and barotropic
approximations.\cite{charney48,hasegawa-maclennan-kodama79,meiss-horton83}
To emphasize its geophysical connections, we shall follow a common 
practice in the plasma physics literature and call the original
form of the Hasegawa--Mima equation the \emph{Charney--Hasegawa--Mima}
(CHM) equation (although in the geophysical literature 
the equation is called the ``quasigeostrophic barotropic potential 
vorticity equation''\cite{lynch03}).

Unfortunately for the interdisciplinarity afforded by the use of the
CHM equation, it became recognized in the early
'90s\cite{dorland_etal90,dorland-hammett93} that a corrected form for
toroidal plasma applications should be used, which we shall call the
\emph{Modified Hasegawa--Mima} (MHM) equation.  Although the
modification seems at first glance to be minor, we shall show that it
makes a profound difference to the modulational stability analysis
because it enhances the generation of zonal flows.

Some early works on modulational instability of drift waves can be
found in Refs.~\refcite{mima-lee80,majumdar88,shivamoggi89} but these
predate the recognition of the need to use the MHM equation to enhance
the nonlinear effect of zonal flows in a toroidal plasma.  Both
Majumdar\cite{majumdar88} and Shivamoggi\cite{shivamoggi89} add a
scalar nonlinearity, arising from polarization drift and/or
temperature gradient, to the CHM equation in order to find a nonlinear
frequency shift.  In the Mima and Lee\cite{mima-lee80} paper, the
nonlinear frequency shift comes from time-averaged flow and density
profile flattening.

In Sec.~(\ref{sec:HM}) we introduce the CHM and MHM equations and in
Sec.~(\ref{sec:NLS}) we introduce the generic form of the nonlinear
Schr\"{o}dinger equation, which describes the time evolution of
modulations on a carrier wave, and use it to derive a criterion for
modulational instability.  In Sec.~(\ref{sec:NLFreq}) we use the MHM
and CHM equations to derive the nonlinear frequency shift of a
finite-amplitude drift/Rossby wave and use it to determine the
criteria for modulational instability of drift and Rossby waves,
respectively.  Section (\ref{sec:Conc}) contains conclusions and
directions for further work.

\section{The CHM and MHM equations}
\label{sec:HM}

The Charney--Hasegawa--Mima equation
(CHM)\cite{hasegawa-maclennan-kodama79,meiss-horton83} is an equation
for the evolution in time, $t$, of the electrostatic potential
$\varphi(x,y,t)$ (or, in the Rossby wave application, the deviation of
the atmospheric depth from the
mean\cite{hasegawa-maclennan-kodama79}).  Here $x$ and $y$ are
Cartesian coordinates describing position in a two-dimensional domain
$\cal D$, representing a cross section of a toroidal plasma with a
strong magnetic field, $\Bvec$, predominantly in the $z$-direction
(unit vector $\zhat$).

In the slab model we take $\cal D$ to be a rectangle with sides of
length $L_x$ and $L_y$.  A circular domain would clearly be more
realistic geometrically because it has a unique central point,
representing the magnetic axis, but it is unlikely to add any
qualitatively new physics.  In strongly shaped tokamaks, like the one
depicted in \fig{fig:JupitervsTokamak}, one might be tempted to give
$\cal D$ the noncircular shape of the plasma edge to add yet more
realism.  However, we caution against this line of thinking because
each point in $\cal D$ represents an \emph{extended section of field
line}, over which the drift-wave amplitude is significant.  That is,
$\cal D$ does not represent any given cross section of the tokamak,
but rather a two-dimensional surface in a field-line coordinate space
(see e.g. Ref.~\refcite{dewar-glasser83}), onto which
behaviour in the third dimension is projected.

We assume the ion temperature to be negligible with respect to the
electron temperature $T_{\rm e}$, assumed constant throughout the
plasma.  The strong magnetic field allows the plasma to support a
cross-field gradient in the time-averaged electron number density,
$\bar{n}$, and the wave dynamics is taken to be sufficiently slow
that, along the field lines, the electrons respond adiabatically to
fluctuations in $\varphi$.  That is, on a given field line they remain
in local thermodynamic equilibrium, with distribution function
$f(\rvec,\vvec,t) = \const \exp(-{\cal E}/T_{\rm e})$, where $T_{\rm
e}$ is the electron temperature in energy units (eV) and $\cal E$ is
the total electron energy $\half mv^2 - e\varphi$, with $m$ the
electron mass and $e$ the electronic charge.

Following Hasegawa and Mima, the shear in the magnetic field is
assumed very weak, so that $z$-derivatives and the parallel component,
$k_{\parallel}$, of the wave vector $\kvec$ can be neglected, thus
reducing the problem to a two-dimensional one.  However, the existence
of magnetic shear is crucial in one qualitative respect---the
foliation of the magnetic field lines into nested toroidal magnetic
surfaces ($x = \const$ in slab geometry).  Field lines cover almost
all magnetic surfaces ergodically, so the constant in the expression
for the distribution function is a surface quantity.  Integrating over
velocity we find
\begin{equation}
    n = n_0(x,t)\exp\left(\frac{e\phitil}{T_{\rm e}} \right)
      = n_0\left(1 + \frac{e\phitil}{T_{\rm e}} + O(\phitil^2) \right) \;,
    \label{eq:nadiab}
\end{equation}
where we have decomposed $\varphi$ into a surface-averaged part,
$\phibar(x,t) \equiv \Pbar\,\varphi(x,y,t)$ (absorbed into $n_0$), and
the surface-varying part, $\phitil \equiv \Ptil\,\varphi \equiv
\varphi - \phibar$.  Here we have used the
\emph{magnetic-surface-averaging operator} $\Pbar$ defined in slab
geometry by
\begin{equation}
    \Pbar\,\cdot \equiv \frac{1}{L_y}\int_{0}^{L_y}\D y\,\cdot \;,
    \label{eq:Pbardef}
\end{equation}
and its complementary projector $\Ptil \equiv 1 - \Pbar$. (Note that 
$\Pbar$ and $\Ptil$ commute with $\partial_t$ and $\grad$.) Equation 
(\ref{eq:nadiab}) can also be derived purely from fluid equations, 
without introducing the distribution function explicitly.

One can show, by surface-averaging the continuity equation for the
electron fluid in the absence of sources or sinks, that the
surface-averaged electron density is independent of $t$.  Thus, to
$O(\phitil)$, $n_0$ is independent of $t$ and equals the prescribed
average density $\bar{n}$.  This would not be the case if we had not
subtracted off $\phibar$ in \eq{eq:nadiab}, and in this we differ from
Hasegawa and Mima but follow most modern practice since Dorland
\emph{et al.}\cite{dorland_etal90,dorland-hammett93} pointed out the
importance of modifying the electron response in this way.  Use of
\eq{eq:nadiab} leads to what we shall call the \emph{Modified
Hasegawa--Mima
equation} (MHM equation). 

Defining a switch parameter $s$ such that $s = 0$ selects the original
CHM equation and $s = 1$ the MHM equation, and a stream function $\psi
\equiv \varphi/B_0$, we write Eq.~(10)
of Hasegawa \emph{et al.}\cite{hasegawa-maclennan-kodama79} as
\begin{equation}
    \frac{\D}{\D t}\left(
    \ln \frac{\omega_{\rm ci}}{n_0} + \frac{\zeta}{\omega_{\rm ci}} - 
    \frac{eB_0}{T_{\rm e}}(\psitil + \delta_{s,0}\psibar)
   \right) = 0 \;,
    \label{eq:Ertel}
\end{equation}
where $\D/\D t \equiv \partial_t + \vvec_E\dotv\grad$, with
\begin{equation}
    \vvec_E
    \equiv
    -\frac{\grad\varphi\cross\zhat}{B_0} = \zhat\cross\grad\psi \;,
    \label{eq:vEdef}
\end{equation}
being the $\Evec\cross\Bvec$ velocity (SI units), $\zeta \equiv
\zhat\dotv\curl\vvec_E = \nabla^2\psi$ the vorticity, $\omega_{\rm ci}
\equiv eB_0/m_{\rm i}$ the ion cyclotron frequency, and $\grad \equiv
\xhat\partial_x + \yhat\partial_y$ the perpendicular gradient.  As
shown in the Appendix of Meiss and Horton\cite{meiss-horton83}, this
is an approximate form of Ertel's theorem for the conservation of
potential vorticity under Lagrangian advection at the
$\Evec\cross\Bvec$ velocity.  Note that the MHM equation satisfies the
expected\footnote{Even in the absence of topography, we do not expect
Galilean invariance in geophysical application of the CHM equation, as
the $\beta$-plane is not an inertial frame.  In the plasma confinement
application, a poloidal boost in polar coordinates would also be to a
rotating frame, but the slab approximation implies we should ignore
any Coriolis effects and Galilean invariance in the poloidal direction
should apply.} Galilean invariance under boosts in the poloidal
direction, $y' = y - Vt$, $\Evec' = \Evec + VB_0\xhat$ (so $\psi' =
\psi - Vx$), whereas the original CHM equation does not and is
therefore unsatisfactory for plasma physics purposes.

We now rewrite \eq{eq:Ertel} in a more explicit way [cf.  Eq.~(1) of
Smolyakov \emph{et al.}\cite{smolyakov-diamond-shevchenko00}]
\begin{equation}
    (\partial_t + \vvec_E\dotv\grad + \vvec_{*}\dotv\grad)
    			(\psitil + \delta_{s,0}\psibar)
    - (\partial_t + \vvec_E\dotv\grad)
			\rhos^2\nabla^2\psi = 0 \;,
    \label{eq:HM}
\end{equation}
where the characteristic drift-wave scale length $\rhos \equiv
\omega^{-1}_{\rm ci}(T_{\rm e}/m_{\rm i})^{1/2}$ is the sound speed
divided by $\omega_{\rm ci}$, and the electron \emph{diamagnetic
drift}\footnote{$v_{*} \equiv | \vvec_{*}|$ is the analogue of $\beta$ in the 
geophysical application of the CHM equation.} is defined by
\begin{equation}
    \vvec_{\rm *} \equiv -\frac{T_{\rm e}\zhat\cross\grad n_0}{eB_0 n_0} \;.
    \label{eq:vdef}
\end{equation}
The ordering in Ref.~\refcite{hasegawa-maclennan-kodama79} makes all
terms in \eq{eq:HM} of the same order.  Thus, balancing
$\partial_t\varphi$ and $\partial_t\rhos^2\nabla^2\varphi^2$ we see
that $\rhos$ is indeed the characteristic scale length for spatial
fluctations.  Balancing $\partial_t\varphi$ and $\vvec_{*}\dotv\grad\varphi$
we see that the characteristic time scale is
$\rhos/v_{\rm *}$, and balancing $v_E$ and $v_{\rm *}$ we see that the
characteristic amplitude of potential fluctuations is $(T_{\rm
e}/e)\rhos/L_n$, where $L_n$ is the scale length for radial variation
of $n_0$.  We assume $\rhos/L_n \ll 1$, so the waves have small
amplitudes compared with the thermal potential.  However, $k\xi$, with
$k$ a typical fluctuation wavelength, and $\xi$ a typical displacement
of a fluid element by the waves, can be order unity, and thus the
equation can describe strong turbulence.

Projecting \eq{eq:HM} with $\Pbar$ and $\Ptil$ we can split it into
two equations, one for the surface-varying part and one for the
zonal-flow part
\begin{eqnarray}
    & (\partial_t + \zhat\cross\grad\psibar\dotv\grad)
      (1-\rhos^2\nabla^2)\psitil
    	+ [\vvec_{\rm *} - 
    	\zhat\cross\grad(\delta_{s,0} 
	-\rhos^2\nabla^2)\psibar]\dotv\grad\psitil & \nonumber \\
    & = \rhos^2\Ptil\zhat\cross\grad\psitil\dotv\grad\nabla^2\psitil \;, &
    \label{eq:HMtil}
\end{eqnarray}
\begin{equation}
    \partial_t(\delta_{s,0} - \rhos^2\nabla^2)\psibar
    = \rhos^2\Pbar\zhat\cross\grad\psitil\dotv\grad\nabla^2\psitil \;.
    \label{eq:HMbar}
\end{equation}
In the MHM case, $s = 1$, \eq{eq:HMbar} reduces to Eq.~(2) of
Ref.~\refcite{smolyakov-diamond-shevchenko00},
$\partial_t\nabla^2\psibar =
-\Pbar\zhat\cross\grad\psitil\dotv\grad\nabla^2\psitil$.

Although physically an approximation, we shall in this paper regard
the CHM/MHM equations as given and treat them as exact equations even
for the mean flow component of $\psi$, which we assume to vary on
longer and slower length and time scales than assumed in the maximal
balance ordering discussed above.

\section{Waves and mean flow}
\label{sec:wave}

\begin{figure}[tbp]
    \centering
     \centerline{\psfig{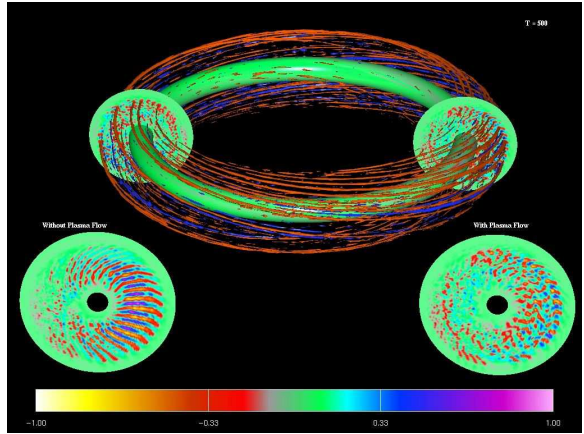}}
    \caption{Visualization showing disruption (lower right) of a
    coherent wave train (lower left) through generation of zonal flows
    by modulational instability in a tokamak simulation.  (Courtesy of
    Z. Lin, \url{http://w3.pppl.gov/~zlin/visualization/}.  See also
    Refs.~\protect\cite{lin_etal98} and
    \protect\cite{chen-lin-white00}.)}
    \label{fig:ZLinFig}
\end{figure}

Assuming there is a scale separation between fluctuations and mean
flow, we introduce an \emph{averaging operation} $\langle\cdot\rangle$
which filters out the fluctuating, wavelike component of whatever it
acts on, leaving only a slowly varying component related to the mean
flow.  This operation can be realized explicitly by convolution with a
smooth, bell-shaped kernel of width (in time and space) long compared
with the fluctuation scale but short compared with the mean flow
scale.  Alternatively we can define it implicitly via the
test-function formalism introduced in Appendix A of
Ref.~\refcite{dewar70}.  Either way, averaging can be shown to commute
with $\partial_t$ and $\grad$ to all orders in $\epsilon$, the ratio
of fluctuation scales to mean-flow scales.

We then split $\psi$ into a slowly varying mean flow part, $\psi_0
\equiv \langle\psi\rangle$, and a fluctuating part, $\psi_1 \equiv 
\psi - \langle\psi\rangle$.

Note that, \emph{except} when the mean flows are purely zonal,
$\langle\cdot\rangle$ is distinct from the surface averaging operation
$\Pbar\cdot$ we used to set up the MHM equation.  In this we differ
from Champeaux and Diamond\cite{champeaux-diamond01}, who, in effect,
take $\Pbar$ to be the same as $\langle\cdot\rangle$ irrespective of
the direction of the mean flows.

We consider the case of modulations carried on a coherent wave (see
e.g. \fig{fig:ZLinFig}), rather than a broad turbulent
spectrum.\cite{krommes-kim00} (As the CHM and MHM equations include no
drift-wave instability mechanism, the origin of this wave is outside
the theory---it is an initial condition.)  Taking, for simplicity,
$v_{*}$ to be a global constant we assume the carrier wave (also
called the pump wave in some analyses) to be a plane wave and write
\begin{equation}
    \psi_1
    = A(\rvec,t)\exp(i\kvec\dotv\rvec - \omega_k t) + {\rm c.c.} \;,
    \label{eq:carrier}
\end{equation}
where $A$ is a slowly varying complex amplitude and c.c. denotes 
complex conjugate.

To begin, we take $A$ and the mean flow, $\langle\vvec_E\rangle$, to
be constant and treat the carrier wave using linear theory.  (Nonlinear
effects will be discussed in Sec.~\ref{sec:NLS}.)  Linearizing
\eq{eq:HM} we find the dispersion relation in the CHM case to be
\begin{equation}
    \omega_k
     = \frac{k_y v_{\rm *}}{1 + \rhos^2 k^2} + 
       \frac{\rhos^2 k^2}
            {1 + \rhos^2 k^2}
       \kvec\dotv\langle\vvec_E\rangle \;,
    \label{eq:disprel0}
\end{equation}
whereas, in the MHM case, $s = 1$, and assuming purely zonal mean flow
($\langle\vvec_E\rangle = \bar{v}_E\yhat$), it is
\begin{equation}
    \omega_k
     = \frac{k_y v_{\rm *}}{1 + \rhos^2 k^2}
     + \kvec\dotv\langle\vvec_E\rangle \;.
    \label{eq:disprel1}
\end{equation}
In the latter case, the mean flow causes a simple Doppler shift of 
frequency, but for the unmodified CHM equation the Doppler shift is 
reduced by a factor $\rhos^2 k^2/(1 + \rhos^2 k^2)$.

We shall use the frequency shift due to mean flow to calculate the
nonlinear frequency shift.  Otherwise we can ignore it.  The
\emph{group velocity}, $\vvec_{\rm g} \equiv
\partial\omega_k/\partial\kvec$, in the absence of a mean flow, is
the same in both cases
\begin{equation}
    \frac{1}{v_{*}}\frac{\partial\omega_k}{\partial\kvec}
     =  \frac{\yhat}{1 + \rhos^2 k^2}
     - \frac{2\rhos^2 \kvec k_y}{(1 + \rhos^2 k^2)^2}\;.
    \label{eq:group}
\end{equation}

We shall also need the dispersion dyadic 
$\grad_{\kvec}\grad_{\kvec}\omega_k$
\begin{equation}
    \frac{1}{v_{\rm *}}\frac{\partial^2\omega_k}{\partial\kvec\partial\kvec}
     = 8\rhos^4 \frac{k_y\kvec\kvec}{(1 + \rhos^2 k^2)^3}
      -2\rhos^2 \frac{\kvec\yhat + \yhat\kvec + k_y{\sf I}}{(1 + \rhos^2 k^2)^2}
     \;,
    \label{eq:dispdyad}
\end{equation}
where $\sf I$ is the unit dyadic.

\section{Nonlinear Schr\"{o}dinger equation and modul\-ational 
instability}
\label{sec:NLS}
We largely follow the simple introduction to modulational instability
theory given in Dewar \emph{et al.},\cite{dewar-kruer-manheimer72}
starting with the \emph{nonlinear Schr\"odinger equation}
\begin{equation}
    \ij\left(\frac{\partial}{\partial t} + 
    \frac{\partial\omega_k}{\partial\kvec}\dotv\grad\right)A
    = \Delta\omega[|A|] A 
    -\frac{1}{2}\frac{\partial^2\omega_k}{\partial\kvec\partial\kvec}
    \ddotv\grad\grad A
    \;,
    \label{eq:NLS}
\end{equation}
where $\partial/\partial\kvec$ denotes the gradient in $\kvec$-space
and $\Delta \omega$ the \emph{nonlinear frequency shift}, a nonlinear
functional of the amplitude $|A|$ (cf.  e.g. Ref.~\refcite{dewar72}).
If the scale length of the modulations is $O(\epsilon^{-1})$ compared
with the wavelength of the carrier, then the $\vvec_{\rm g}\dotv\grad$
term on the LHS of \eq{eq:NLS} is $O(\epsilon)$ whereas the
$\grad_{\kvec}\vvec_{\rm g}\ddotv\grad\grad$ term on the RHS is
smaller, $O(\epsilon^2)$.  Assuming the nonlinear frequency shift to
be of the same order, we see that \eq{eq:NLS} expresses the fact that,
on a short time scale, modulations simply advect with the group
velocity, while on a longer timescale the nonlinear frequency shift
causes a slow drift in the phase while the dispersion dyadic
$\partial^2\omega_k/\partial\kvec\partial\kvec$ causes spreading of
the modulations.

An amplitude-modulated wave can be represented as the sum of the
unmodulated carrier wave and upper and lower sidebands
\begin{eqnarray}
    A & = & A_0\exp(-\ij\Delta\omega_0 t) \nonumber\\
      &   & \times
        \left\{
	  1
	  + a_{+}\exp(\ij\Kvec\dotv\rvec - \ij\Omega t)
	  + a_{-}^{*}\exp(-\ij\Kvec\dotv\rvec + \ij\Omega^{*} t)
	\right\}
    \;,
    \label{eq:modulate}
\end{eqnarray}
where $\Delta\omega_0 \equiv \Delta\omega[|A_0|]$.

Linearizing in $|a_{\pm}|$, $|A| = |A_0|[1 + \half(a_{+} + 
a_{-})\exp \ij(\Kvec\dotv\rvec - \Omega t) + {\rm c.c.}]$, and using 
this in \eq{eq:NLS} we find
\begin{eqnarray}
    & & \hspace{-4mm}
    \left[
    \begin{array}{cc}
        \Omega - \Kvec\dotv\frac{\partial\omega_k}{\partial\kvec}
	-\frac{\Kvec\Kvec}{2}{\rm :}\frac{\partial^2\omega_k}{\partial\kvec\partial\kvec}
	-\frac{1}{2}\delta\omega_K &
        -\frac{1}{2}\delta\omega_K	\\
	\frac{1}{2}\delta\omega_K &
	\Omega - \Kvec\dotv\frac{\partial\omega_k}{\partial\kvec}
	+\frac{\Kvec\Kvec}{2}{\rm :}\frac{\partial^2\omega_k}{\partial\kvec\partial\kvec}
	+\frac{1}{2}\delta\omega_K
    \end{array}
    \right]
    \left[
    \begin{array}{c}
        a_{+}  \\
        a_{-}
    \end{array}
    \right] \nonumber\\
    & & = 0 \;,
    \label{eq:matrix}
\end{eqnarray}
where $\delta\omega_K$ (denoted $\alpha\Delta\omega_0$ in
Ref.~\refcite{dewar-kruer-manheimer72}) is defined by
\begin{equation}
    \delta\omega_K \equiv |A_0|\exp(-\ij\Kvec\dotv\rvec)
    \int\D^2 x 
    \frac{\delta\Delta\omega}{\delta |A|} \exp(\ij\Kvec\dotv\rvec)
    \;.
    \label{eq:alphadef}
\end{equation}

Setting the determinant of the matrix in \eq{eq:matrix} to zero gives
the dispersion relation for plane-wave modulations
\begin{equation}
    \left(\Omega - 
    \Kvec\dotv\frac{\partial\omega_k}{\partial\kvec}\right)^2
    = \frac{1}{2}\Kvec\Kvec{\rm :}
    \frac{\partial^2\omega_k}{\partial\kvec\partial\kvec}
    \left(
    \delta\omega_K + 
    \frac{1}{2}\Kvec\Kvec{\rm :}
	\frac{\partial^2\omega_k}{\partial\kvec\partial\kvec}
    \right)
    \;.
    \label{eq:moddisprel}
\end{equation}
The criterion for modulational instability is that $\Omega$ be
complex, $\Omega = \Omega_{\rm r} + \ij\Gamma$, $\Gamma > 0$, and from
\eq{eq:moddisprel} we immediately see that this occurs, for 
sufficiently small $K$, if and only if there exist directions for 
$\Kvec$ in which
\begin{equation}
    \delta\omega_K 
    \Kvec\Kvec{\rm :}
	    \frac{\partial^2\omega_k}{\partial\kvec\partial\kvec}
	    < 0 \;.
    \label{eq:instabcrit}
\end{equation}

\section{Nonlinear frequency shift}
\label{sec:NLFreq}

The nonlinear frequency shift $\Delta\omega$ in a general fluid or
plasma is composed of two parts.  The first is that due to the
intrinsic nonlinearity of the medium and the second is that due to
Doppler-like shifts [see Eqs.~(\ref{eq:disprel0}) and
(\ref{eq:disprel1})] associated with nonlinearly induced mean flows.

However, in the case of drift or Rossby waves described by the CHM or
MHM equations, \emph{the intrinsic nonlinear frequency shift is zero}
(or, at most, of higher order than quadratic).  To see this, consider
the terms in \eq{eq:HM} describing nonlinear wave-wave (including
self) interactions: $\{\psi_1,\psitil_1+\delta_{s,0}\psibar_1 \}$ and
$\{\psi_1,\nabla^2\psi_1\}$, where the \emph{Poisson bracket} of two
functions $f$ and $g$ is defined by
\begin{equation}
    \{f,g\}
    \equiv \zhat\cross\grad f\dotv\grad g
    = \partial_x f.\partial_y g - \partial_x g.\partial_y f \;.
    \label{eq:poissondef}
\end{equation}
Since we assume $k_y \neq 0$, $\psibar_1$ is zero (to leading order,
at least) and $\psitil_1 = \psi_1$.  As the Poisson bracket $\{f,f\}
\equiv 0$ for any $f$, the first nonlinear self-interaction term
vanishes.  Similarly, because we are considering a monochromatic
carrier wave, the second self-interaction term also vanishes to 
leading order:
$\{\psi_1,\nabla^2\psi_1\} \approx -k^2\{\psi_1,\psi_1\} \equiv 0$.

For the calculation of the nonlinearly excited mean flows, we will 
need to evaluate the above term more accurately, which is best done 
via the useful identity
\begin{equation}
    \{f,\nabla^2 f \}
    = \partial_x\partial_y
    \left[(\partial_x f)^2 - (\partial_y f)^2\right]
    - \left(\partial_x^2  - \partial_y^2\right)
    (\partial_x f.\partial_y f) \;.
    \label{eq:poissonident}
\end{equation}
(Some earlier discussion of this identity
can be found in Ref.~\refcite{krommes04}.)
Averaging Eqs.~(\ref{eq:HMtil}) and (\ref{eq:HMbar}) over the 
fluctuation scale,
\begin{eqnarray}
    (\partial_t + \vvec_{*}\dotv\grad)
      (1-\rhos^2\nabla^2)\langle\psitil\rangle
     & = & \rhos^2\Ptil\langle\{\psitil_1,\nabla^2\psitil_1\}\rangle \;,
    \label{eq:HMtilav} \\
    \partial_t(\delta_{s,0} - \rhos^2\nabla^2)\langle\psibar\rangle
    & = & \rhos^2\Pbar\langle\{\psitil_1,\nabla^2\psitil_1\}\rangle \;.
    \label{eq:HMbarav}
\end{eqnarray}

For both the CHM and MHM cases, the small $\rhos^2\nabla^2 = O(\rhos^2
K^2)$ term on the LHS in \eq{eq:HMtilav} is negligible compared with
1.

\subsection{Modulational instability for MHM equation}
\label{sec:modstabMHM}

However, on the LHS of \eq{eq:HMbarav} the leading term 1 does not
occur in the MHM case, $s = 1$, so the $\rhos^2\nabla^2$ term must be
retained.  Consequently, in this case $\langle\psitil\rangle$ is
smaller than $\langle\psibar\rangle$ by a factor $O(\rhos^2 K^2)$.
That is, the mean flow, $\propto\zhat\cross\Kvec$, is predominantly
zonal so we lose no real generality in assuming $\Kvec = K\xhat$ and
setting $\langle\psitil\rangle = 0$.  Then, using the identity
\eq{eq:poissonident}, dividing \eq{eq:HMbarav} by $\rhos^2$ and
integrating twice with respect to $x$ we
find\cite{smolyakov-diamond-shevchenko00}
\begin{equation}
    \partial_t\psi_0(x,t)
    = \Pbar\partial_x\psi_1.\partial_y\psi_1
    = 2k_x k_y |A|^2\;,
    \label{eq:HMbaravint1}
\end{equation}
with the second form following from \eq{eq:carrier}.

We can convert the time derivative to a spatial derivative by noting 
that the RHS of \eq{eq:NLS} is small (assuming $|A|$ is small) so, 
to leading order, $\partial_t A = -\vvec_{\rm g}\dotv\grad A$; that 
is, the modulations move at the group velocity. This 
also applies to quantities like $\psi_0$ driven by $|A|$, so, to 
leading order, \eq{eq:HMbaravint1} becomes
\begin{equation}
    \partial_x\psi_0
    = -\frac{2k_x k_y}{v_{{\rm g}x}} |A|^2\;.
    \label{eq:HMmeanflo1}
\end{equation}
By \eq{eq:vEdef}, $\overline{\vvec}_{E} = \partial_x\psi_0\yhat$, so
Eqs.~(\ref{eq:disprel1}) and (\ref{eq:group}) give the nonlinear
frequency shift for case $s = 1$ as a simple function of $|A|$
\begin{equation}
    \Delta\omega
    = \frac{(1 + \rhos^2 k^2)^2 k_y}{\rhos^2 v_{\rm *}} |A|^2\;.
    \label{eq:NLfreqMHM}
\end{equation}

As the nonlinear frequency shift in this case is a simple quadratic
function (rather than functional) of $|A|$, the modulated frequency
shift parameter $\delta\omega_K$, \eq{eq:alphadef}, in the nonlinear
Schr\"odinger equation is now given simply by $\delta\omega_K =
\D\Delta\omega/\D\ln |A| = 2\Delta\omega_0$.  The modulational
instability criterion, \eq{eq:instabcrit}, now becomes
$-k_y\partial^2\omega_k/\partial k_x^2 < 0$.  That is, from
\eq{eq:dispdyad}, the modulational instability criterion for the
modified Hasegawa--Mima equation case, $s = 1$, is
\begin{equation}
     1 - 3\rhos^2 k_x^2 + \rhos^2 k_y^2 > 0 \;,
    \label{eq:modstabcrit1}
\end{equation}
which agrees with Ref.~\refcite{smolyakov-diamond-shevchenko00}
and Ref.~\refcite{krommes06} 
but not with Ref.~\refcite{champeaux-diamond01} who,
due to a misprint\cite{diamond_prcom06} reproduced in
Ref.~\refcite{diamond-itoh-itoh-hahm05}, omit the factor 3 
multiplying $\rhos^2 k_x^2$. (An apparently similar inequality to that 
in Ref.~\refcite{champeaux-diamond01} appears in
Ref.~\refcite{guzdar-kleva-chen01}, but this is not really relevant 
as they consider only a drift wave propagating in the poloidal 
direction---their $k_x$ is our $K_x$.)

If criterion \eq{eq:modstabcrit1} is fulfilled,
the growth rate curve, $\Gamma^2$ \emph{vs.} $K^2$, is an inverted
parabola with maximum at
\begin{equation}
    \Gamma_{\rm max} = \Delta\omega_0, \quad
    K_{\rm max} = \frac{(1+\rhos^2 k^2)^{5/2}}{(1 - 3\rho^2 k_x^2 + 
    \rhos^2 k_y^2)^{1/2}}\frac{|A|}{\rhos^2 v_{\rm *}} \;.
    \label{eq:gammaKmax}
\end{equation}
This extends the small-$K$ result in Eq.~(15) of
Ref.~\refcite{smolyakov-diamond-shevchenko00} (who use the notation
$q$ for our $K$) to get a turnover in $\Gamma$ at large $K$, as was
also found using a mode-coupling approach by Chen \emph{et
al.}\cite{chen-lin-white00} via the gyrokinetic equation in toroidal
geometry within the ballooning approximation, and by Lashmore-Davies
\emph{et al.}\cite{lashmored-mccarthy-thyagaraja01} using the 
modified Hasegawa--Mima equation.

As the latter authors base their analysis on the same model as used in
the present paper, we can make a precise comparison between our
\eq{eq:moddisprel} and their modulational dispersion relation,
Eq.~(43) in the small $q \equiv K_x$ limit implied by our envelope
approach.  Expanding their quantities $\delta_{\pm}$ in $q$, it is
easily seen that, to leading order, their $\delta_{+} + \delta_{-} =
q^2\partial^2\omega_k/\partial k_x^2$ and $(\delta_{+} - \delta_{-})/2
= q\partial\omega_k/\partial k_x$, while their expression $2\Omega_0^2
|A_0|^2 F_0(k_0,q)$ is $\delta\omega_K = 2\Delta\omega_0$, with
$\Delta\omega_0$ given by \eq{eq:NLfreqMHM} above.  Completing the
square in their Eq.~(43), we see that correspondence between the two
modulational dispersion relations is exact in the small $q = K_x$
limit.

Note also that the modulations and zonal flows have finite frequency,
even without geodesic effects,\cite{winsor-johnson-dawson68} as they
propagate radially with phase velocity equal to the carrier group
velocity, $\partial\omega_k/\partial k_x = - 2\rhos^2 k_x k_y v_{\rm
*}/(1 + \rhos^2 k^2)^2$ from \eq{eq:group}.

\subsection{Modulational instability for CHM equation}
\label{sec:modstabCHM}

In the unmodified case, $s = 0$, there is actually no compelling
reason to make the split into zonal and nonzonal components.
Averaging \eq{eq:HM} over the fluctuation scale (noting that
$\vvec_E\dotv\grad\psi \equiv 0$ and neglecting small terms) we find
\begin{equation}
    (\partial_t + \vvec_{\rm *}\dotv\grad)\psi_0
    = \rhos^2\langle
    		\vvec_E\dotv\grad\nabla^2\psi_1
	     \rangle
    = \rhos^2\langle
    		\{\psi_1,\nabla^2\psi_1\}
	     \rangle
    \;.
    \label{eq:CHMaveraged}
\end{equation}
Integrating from $t = -\infty$, where $\psi_0$ is assumed to vanish,
along the trajectory of a fluid element moving at the drift speed and
assuming the modulations in the forcing term on the RHS to be
advecting at the group velocity we find
\begin{equation}
	\psi_0(x,y,t)
	= \rhos^2\int_{-\infty}^{0}\D \tau
	 \langle \{\psi_1,\nabla^2\psi_1\}
	 \rangle
	 \left(
	    x - \frac{\partial\omega_k}{\partial k_x}\tau,
	    y + (v_{\rm *} - \frac{\partial\omega_k}{\partial k_y})\tau,
	    t
	 \right)
\;.
\label{eq:CHMavint}
\end{equation}

Using the ansatz \eq{eq:carrier}, the dispersion relation
\eq{eq:disprel0}, and the identity \eq{eq:poissonident} we find the 
nonlinear frequency shift to be the functional
\begin{eqnarray}
    \Delta\omega & = & \
    \frac{2\rhos^4 k^2}{1 + \rhos^2 k^2}
    (k_y\partial_x - k_x\partial_y)
    [(k_x^2 - k_y^2)\partial_x\partial_y
    - k_x k_y(\partial_x^2 - \partial_y^2)]
\nonumber \\  & & \mbox{}
    \times\int_{-\infty}^{0}\D\tau|A|^2
    \left(
       x - v_{{\rm g}x}\tau,
       y + (v_{*} - v_{{\rm g}y})\tau,
       t
    \right)
\;.
    \label{eq:NLfreqCHM}
\end{eqnarray}

Perturbing $\Delta\omega$ with a small modulation $\delta|A|$,
replacing $\delta|A|$ by $\exp \ij\Kvec\dotv\rvec$, and substituting 
in \eq{eq:alphadef} we get the modulated frequency shift
\begin{equation}
	\delta\omega_K = 
	\frac{4\rhos^4 k^2|A_0|^2}{1 + \rhos^2 k^2}
	\frac{(k_y K_x - k_x K_y)
	    [(k_x^2 - k_y^2) K_x K_y
	    - k_x k_y(K_x^2 - K_y^2)]}
	    {K_x v_{{\rm g}x} + K_y(v_{{\rm g}y} - v_{\rm *})}
\;.
\label{eq:delomK_CHM}
\end{equation}
Equation~(\ref{eq:moddisprel}) then gives the dispersion relation for
small modulations.  Clearly, this is considerably more complicated
than found in the MHM case and will not be analyzed further here
except to make comparison with the results of
Ref.~\refcite{smolyakov-diamond-shevchenko00}, who take $K_y = 0$.  In
this case
\begin{equation}
	\delta\omega_K = 
	\frac{2|A_0|^2 K_x^2}{v_{\rm *}}\rhos^2 k^2(1 + \rhos^2 k^2)
\;,
\label{eq:delomK_CHMzonal}
\end{equation}
which gives a modulational dispersion relation in the small $K_x$
limit in essential agreement with Eq.~(19) of
Ref.~\refcite{smolyakov-diamond-shevchenko00}, who note that the
modulational instability criterion is the same as that for the MHM,
\eq{eq:modstabcrit1}.  However, the resonance at
$\Kvec\dotv(\vvec_{\rm g} - \vvec_{\rm *}) = 0$ arising from the
vanishing of the denominator in \eq{eq:delomK_CHM} will give higher
growth rates for oblique modulations, so it is not clear that this
special case is of great significance for Rossby waves in the absence
of boundaries.

\section{Conclusions}
\label{sec:Conc}

We have derived a nonlinear Schr\"odinger equation for modulations on
a train of drift or Rossby waves in a very universal, if heuristic,
fashion.  The nonlinear Schr\"odinger equation has been widely studied
in other applications and is known to have soliton solutions.
However, we have analyzed it only for stability to small modulations
and have found criteria in agreement with those found by Smolyakov
\emph{et al.}\cite{smolyakov-diamond-shevchenko00} for modulation
waves with zonal phase fronts.

Our results are encouraging as a step towards explaining the
experimental discovery by Shats and Solomon\cite{shats-solomon02} of
modulational instability associated with low-frequency zonal flows,
but the Hasegawa--Mima equation is rather too simplified for direct
comparison with experiment and further work remains to be done 
in this regard.

\section{Acknowledgments}
This work was supported by the Australian Research Council and AusAID.
We thank Dr F.L. Waelbroeck for explaining the importance of the 
modification of the electron adiabatic response leading to the Modified 
Hasegawa--Mima Equation and Dr G.W. Hammett for discussions on the 
history of this modification. Also we thank the referee for 
constructive suggestions and Dr J.S. Frederiksen for commenting on 
the nomenclature difference between the plasma and geophysical 
communities regarding the CHM equation, and Dr R. Ball for bringing 
Ref.~\refcite{lynch03} to our attention.

\bibliography{turbulence}

\end{document}